\documentclass[doublecol]{epl2} 

\usepackage{dcolumn}
\usepackage{amssymb,amsmath,amsthm}
\usepackage{graphicx}
\usepackage{epsf,epsfig}

\usepackage{color}

\newcommand{\p}[1]{\left(#1\right)}
\newcommand{\mE}{\mathcal{E}}
\newcommand{\mP}{\mathcal{P}}
\newcommand{\mR}{\mathcal{R}}
\newcommand{\mA}{\mathcal{A}}
\newcommand{\Lf}{\mathcal{L}}
\newcommand{\tr}{\mathsf{tr}}

\newcommand{\comment}[1]{}

\title{Statistics of sums of correlated variables described by a matrix product ansatz}

\author{Florian Angeletti$^{1,2,3}$, Eric Bertin$^{1,4}$, and Patrice Abry$^1$}
\institute{$^1$ Universit\'e de Lyon, Laboratoire de Physique, ENS Lyon, CNRS,
46 All\'ee d'Italie, F-69007 Lyon, France\\
$^2$ National Institute for Theoretical Physics (NITheP), Stellenbosch 7600, South Africa\\
$^3$ Institute of Theoretical Physics, University of Stellenbosch, Stellenbosch 7600, South Africa\\
$^4$ Laboratoire Interdisciplinaire de Physique, Universit\'e Joseph Fourier Grenoble, CNRS UMR 5588, BP 87, F-38402 Saint-Martin d'H\`eres, France\\
}

\date{\today}


\abstract{We determine the asymptotic distribution of the sum of correlated variables described by a matrix product ansatz with finite matrices, considering variables with finite variances. In cases when the correlation length is finite, the law of large numbers is obeyed, and the rescaled sum converges to a Gaussian distribution. In contrast, when correlation extends over system size, we observe either a breaking of the law of large numbers, with the onset of giant fluctuations, or a generalization of the central limit theorem with a family of nonstandard limit distributions. The corresponding distributions are found as mixtures of delta functions for the generalized law of large numbers, and as mixtures of Gaussian distributions for the generalized central limit theorem. Connections with statistical physics models are emphasized.}

\begin{document}

\maketitle

\section{Introduction}

The law of large numbers and the central limit theorem are cornerstones of equilibrium statistical physics. Indeed, the very existence of deterministic values
of macroscopic observables in large systems relies on the law of large numbers, while the Gaussian shape of tiny fluctuations around the mean value are described by the central limit theorem.
The latter also bears strong connections with random walks \cite{FellerI-66}.
Basic forms of these theorems are known for independent
and identically distributed (i.i.d.) random variables
\cite{Gnedenko54,FellerI-66,FellerII-66}.
Some generalizations with less restrictive assumptions are also known
\cite{FellerII-66,Petrov95}, and it is often assumed that these theorems remain valid as long as
distributions of individual variables are not too broad and do not
differ too much one from the other,
and as long as correlations are weak enough.
These assumptions are however not always valid in nonequilibrium systems,
where the relevant observables may have an infinite mean
(e.g., in aging \cite{Bouchaud90,Bouchaud92} or laser cooling \cite{Bardou02} phenomena),
or may have long-range correlations (e.g., in boundary driven \cite{DerridaEvans93,Essler96} or active \cite{Ramaswamy03,Toner05} systems), leading to the breakdown of the law of large numbers.
Similarly, the standard central limit theorem breaks down in a number of cases. For broadly distributed variables with infinite variance, the generalized central limit theorem yields L\'evy-stable laws \cite{Gnedenko54}, with many applications often related to anomalous diffusion \cite{Bouchaud90}.
Non-Gaussian distributions have also been found for instance in the context
of $1/f^{\alpha}$-noise \cite{Antal02} and related problems \cite{Bramwell01,Clusel08}, where summed variables have very different statistics.
For strongly correlated variables, generalizations of the central limit theorem have been derived for Gaussian processes \cite{Taqqu79,Rosenblatt81}.
However, in a statistical physics context, another class of strongly correlated variables, defined through a matrix product ansatz, has emerged from the exact solution of many types of one-dimensional nonequilibrium models,
like the Asymmetric Simple Exclusion Process (ASEP) \cite{Hakim83,DerridaASEP93,Essler96,Mallick97,Evans07,Crampe2011,Lazarescu2011,Lazarescu2013} and generalizations including several types of particles \cite{Mallick09,Evans07},
as well as different types of reaction-diffusion processes \cite{Hinrichsen96,Hinrichsen00,Hieida04,Evans07,Basu09,Hinrichsen13}, including e.g.,
the Branching-Coalescing Random Walk and the Asymmetric Glauber-Kawasaki Process \cite{Jafarpour03,Jafarpour04}.
Note that such types of lattice models have proven useful in the study of intracellular motility \cite{Frey03} and of vehicular traffic \cite{Franosch07,Appert10}.
The matrix product ansatz has also been used recently to solve coupled KPZ equations \cite{Spohn13}.
Although infinite matrices may be needed, notably in the context
of the ASEP model \cite{DerridaASEP93,Evans07}, many models can be solved using finite matrices, including reaction-diffusion models \cite{Jafarpour03,Jafarpour04,Hinrichsen96,Hinrichsen00,Hieida04,Evans07,Basu09,Hinrichsen13}, the coupled KPZ equations \cite{Spohn13} and the ASEP model for specific parameter values \cite{Essler96,Speer97,Mallick97}.
In spite of the increasing importance of this class of random variables in the description of correlated non-equilibrium systems, the corresponding generalizations of the law of large numbers and of the central limit theorem are presently not known.
In this Letter, we aim at providing such generalizations for variables described by a matrix product ansatz with finite matrices.
Our study encompasses both discrete random variables as in the ASEP model and reaction-diffusion processes \cite{DerridaASEP93,Evans07}, and continuous ones as in signal processing \cite{Angeletti-ICASSP,Angeletti-TSP} or in mass transport models \cite{Bertin-unpublished}.
For the sake of clarity, we restrict our presentation to specific, yet representative cases, deferring a full-length account of our results to a forthcoming publication \cite{Angeletti-unpublished}.

\section{Matrix product representation}

Following \cite{Angeletti-ICASSP,Angeletti-TSP},
we consider a set of random variables $(x_1,\ldots,x_N)$ whose joint
probability distribution can be described by a matrix product ansatz, namely
\begin{equation}
\label{eq:Def}
P(x_1,\dots,x_N) = \frac{1}{\Lf (\mE^N)}\, \Lf \big( \mR(x_1) \mR(x_2) \ldots \mR(x_N) \big)
\end{equation}
where $\mR(x)$ is a $D \times D$ matrix function ($D \ge 2$) with real
nonnegative entries, $\mE=\int_{-\infty}^{\infty} \mR(x)\,dx$,
and $\Lf$ is a linear form defined as
\begin{equation}
\label{eq:def-L}
\Lf\p{M} = \tr\p{\mA^T M},
\end{equation}
with $\mA$ a given $D \times D$, nonzero matrix with real nonnegative
entries \color{black}{(an extension relaxing this positivity condition for $\mA$ will be mentioned at the end of this letter)}.
We further assume that for all $N \ge 1$, $\Lf(\mE^N) \ne 0$.
Standard forms used in statistical physics for the linear form $\Lf$
\cite{Evans07}
are recovered either by taking $\mA$ as the identity matrix, or by choosing
$\mA_{ij}=V_i W_j$ so that $\Lf(M) = \langle V|M|W\rangle$.
Eq.~\eqref{eq:Def} is a natural generalization
to correlated variables of the i.i.d.~case, replacing the product of
real functions by a product of matrix functions.
It is useful to introduce the matrix
of distributions $\mP(x)$ through the relation
\begin{equation}
\mR_{ij}(x) = \mE_{ij} \mP_{ij}(x),
\end{equation}
so that $\mP_{ij}(x)$ can be interpreted as a probability distribution,
normalized to $1$. \color{black}{The definition~\eqref{eq:Def} is valid for any probability space, however in the present letter,
we restrict ourselves to real random variables. Moreover,}
we consider probabilities $\mP_{ij}(x)$ with finite mean value $m_{ij}$
and finite variance $\sigma_{ij}^2$.
Note that $\mP_{ij}(x)$ is uniquely defined only when $\mE_{ij} \neq 0$.

\color{black}{ Using the product structure of Eq.~\eqref{eq:Def}, the correlation can be computed as
\begin{equation} \label{eq:correlation}  \begin{gathered} 
C_{kl} \equiv \langle x_k x_l \rangle - \langle x_k \rangle \langle x_l \rangle \\
\langle x_k \rangle = \frac{ \Lf (\mE^{k-1} M(1) \mE^{N-k-1}) } {\Lf(\mE^N)} \\
\langle x_k x_l \rangle =  \frac{ \Lf (\mE^{k-1} M(1) \mE^{l-k-1} M(1) \mE^{N-l}) } {\Lf{(\mE^N)}} 
\end{gathered} \end{equation}
where $k<l$ and $M(q)$  is the moment matrix $M(q) = \int x^q \mR(x) dx$.
This expression of the correlation will be useful in the following.
}

\section{Hidden Markov Chain representation}

As shown in \cite{Angeletti-ICASSP,Angeletti-TSP}, the joint probability
(\ref{eq:Def}) can be reinterpreted using the concept of Hidden Markov Chain
\cite{Cappe05}. We introduce a Markov chain $\Gamma\color{black}{ \in \{1,\dots D\}^{N+1}}$ such
that
\begin{align}
\label{eqn:trans0}
{\rm Pr}(\Gamma_1 = i,\Gamma_{N+1} = f) &= \mA_{if} \frac{(\mE^N)_{if}}{\Lf(\mE^N)} \; , \\
 \label{eq:transition}
{\rm Pr}(\Gamma_{k+1}= j | \Gamma_{k}=i, \, \Gamma_{N+1} = f ) &= \mE_{ij} \frac{(\mE^{N-k})_{jf}}{(\mE^{N-k+1})_{if}} \; .
\end{align}
Note that this Markov chain is non-homogeneous and of a nonstandard type, due to the dependence on the final state $\Gamma_{N+1}$.
\color{black}{In particular for $k=N$, the transition rate  ${\rm Pr}(\Gamma_{k+1}= j | \Gamma_{k}=i, \, \Gamma_{N+1} = f )$
equals $1$ if $j=f$ and $0$ otherwise.}
\color{black}{Combining Eqs~\eqref{eqn:trans0} and \eqref{eq:transition}, the global probability of a given chain $\Gamma$ reads}
\begin{equation} \label{eq:kappa}
\color{black}{ \kappa_\Gamma = \frac{\mA_{\Gamma_1 \Gamma_{N+1}}}{\Lf\p{\mE^N}}\, \mE_{\Gamma_1 \Gamma_2} \dots \mE_{\Gamma_N \Gamma_{N+1}} }
\end{equation}
For a given $\Gamma$, the random variables $(x_1,\ldots,x_N)$ are independent but non-identically distributed, 
with a probability distribution depending on $\Gamma$:
\begin{equation}\label{eq:X|gamma}
P_{\Gamma}(x_1,\ldots,x_N) = \prod_{k=1}^N \mP_{\Gamma_k \Gamma_{k+1}}(x_k) \;.
\end{equation}
This formulation using a hidden Markov chain $\Gamma$ is equivalent
to the definition Eq.~\eqref{eq:Def} using matrices \cite{Angeletti-ICASSP,Angeletti-TSP}.
This yields a procedure to simulate the correlated random variables described
by Eq.~\eqref{eq:Def} \cite{Angeletti-TSP}: (i) $\Gamma_1$ and $\Gamma_{N+1}$
are chosen at random according to distribution \eqref{eqn:trans0};
(ii) the random chain $\Gamma$ is obtained from transition rates
\eqref{eq:transition}; (iii) the random variables $x_k$ ($k=1,\ldots N$)
are drawn randomly from the distributions $\mP_{\Gamma_k \Gamma_{k+1}}(x_k)$.

As seen in Eq.~\eqref{eq:X|gamma}, for a fixed $\Gamma$, the random variables
$(x_1,\ldots,x_N)$ are independent.
Correlations, when present, thus emerge from the correlation of the hidden chain $\Gamma$. Consequently, the matrix $\mE$ plays a key role
in statistical properties of the Markov chain $\Gamma$, which in turn
determine correlations between the $x_k$'s.
\color{black}{In particular, the short-range or long-range nature of the correlations depends respectively on the ergodic or non-ergodic nature of the Markov chain $\Gamma$. This ergodic nature is determined by the matrix $\mE$ (see examples in \cite{Angeletti-TSP} where $\mE$ is either diagonalizable or given by the sum of the identity and a nilpotent matrix).} This characteristic of the chain $\Gamma$ is key to the classification of the representative examples used below.

\section{Statistics of the sum}

We now focus on the study of the statistical properties of the sum
$S=\sum_{k=1}^N x_k$, in the limit $N\to \infty$.
We are specifically interested in the validity
of the law of large numbers and of the central limit theorem.
Introducing the variable $s=S/N$, the law of large numbers breaks down if
the limit distribution $\Psi(s)$ for $N \to \infty$ does not reduce to a delta function.
When the law of large numbers holds, namely $\Psi(s)=\delta(s-m)$,
one can investigate the fluctuations of $S$ around its mean value $Nm$
on a scale $\sim \sqrt{N}$, through the rescaled variable
\begin{equation}
\label{eq:def-z}
z = \frac{S-Nm}{\sqrt{N}}.
\end{equation}
When correlations are weak enough, the central limit theorem should hold,
and the distribution $\Phi_N(z)$ should converge to a Gaussian
when $N \to \infty$.

The Hidden Markov Chain formalism proves very useful in order to study
the statistical properties of the sum, as it allows one to separate two different sources of randomness: the random choice of the chain $\Gamma$,
and the random choice of $(x_1,\ldots,x_N)$ from the distribution
$P_{\Gamma}$, for a fixed $\Gamma$.
Hence the distribution of $S$ can be determined by
first computing the distribution of $S_{\Gamma}=\sum_{k=1}^N x_k$, where
the $x_k$'s are drawn from $P_{\Gamma}$, and then averaging
the distribution of $S_{\Gamma}$ over $\Gamma$.
For a given chain $\Gamma$, we introduce the fraction $\nu_{ij}$ of transitions
from $i$ to $j$ in $\Gamma$,
\begin{equation}
\nu_{ij} = \frac{1}{N}\, {\rm card}\{k, \Gamma_k=i \; {\rm and} \;
\Gamma_{k+1}=j\}\,.
\end{equation}
For a given $\Gamma$, the sum $S_{\Gamma}$ can be rewritten as
\begin{equation} \label{eq:sum:rew}
S_{\Gamma} = \sum_{i,j=1}^D \sum_{k=1}^{N \nu_{ij}} X_k^{(ij)},
\end{equation}
where the variables $X_k^{(ij)}$, $k=1,\ldots,N\nu_{ij}$, are i.i.d.~random
variables drawn from the distribution $\mP_{ij}$.
Let us first investigate the validity of the law of large numbers
for the sum $S$. For a fixed $\Gamma$, one can write
\begin{equation}
\frac{S_{\Gamma}}{N} = \sum_{i,j=1}^D \nu_{ij} \left(
\frac{1}{N \nu_{ij}} \sum_{k=1}^{N \nu_{ij}} X_k^{(ij)} \right).
\end{equation}
Given that, for fixed $(i,j)$, the variables $X_k^{(ij)}$ are i.i.d.,
the law of large numbers can be applied to the term between brackets,
which thus converges (almost surely) to the mean value $m_{ij}$.
Hence, for a fixed $\nu=(\nu_{ij})$, the conditional distribution of $s$
converges when $N\to\infty$ to
\begin{equation}
\Psi(s|\nu) = \delta\left( s-\sum_{i,j=1}^D \nu_{ij} m_{ij} \right).
\end{equation}
As the only dependence over $\Gamma$ is through $\nu$,
the average over $\Gamma$ can be replaced by an average over $\nu$.
The limit distribution $\Psi(s)$ is thus obtained by averaging
$\Psi(s|\nu)$ over the asymptotic ($N\to \infty$) distribution
of $\nu$, denoted as $Q(\nu)$:
\begin{equation}
\label{eq:Psi-s}
\Psi(s) = \int \prod_{i,j=1}^D d\nu_{ij}\, Q(\nu) \;
\delta \!\! \left( s-\sum_{i,j=1}^D \nu_{ij} m_{ij} \right).
\end{equation}
The law of large numbers holds either when all $m_{ij}$'s are equal
(or at least those associated to nonzero $\nu_{ij}$), or
when the empirical frequencies $\nu_{ij}$ converge to nonrandom values
$\overline{\nu}_{ij}$ in the limit $N\to \infty$, in which case
the rescaled sum $s$ converges to the deterministic limit
$m=\sum_{i,j} \overline{\nu}_{ij} m_{ij}$.
When the law of large numbers is satisfied, the validity of the
central limit theorem can then be investigated.
Given that $m$ is non-random, we use the variable $z$
defined in Eq.~\eqref{eq:def-z}, and follow a similar path as above.
For a given chain $\Gamma$, and thus a given $\nu$, one has
\begin{equation}
\frac{S_{\Gamma}-Nm}{\sqrt{N}} = \sum_{i,j=1}^D \sqrt{\nu_{ij}} \left(
\frac{\sum_{k=1}^{N \nu_{ij}} (X_k^{(ij)} -m_{ij})}{\sqrt{N\nu_{ij}}}\right).
\end{equation}
As for fixed $(i,j)$, the variables $X_k^{(ij)}$ are i.i.d.~with finite
variance, the central limit theorem applies to the sum between
brackets, and the distribution of this sum converges, for $N\to \infty$,
to a centered Gaussian distribution of variance $\sigma_{ij}$.
Hence the conditional distribution $\Phi(z|\nu)$ is a centered
Gaussian distribution of variance $\sum_{i,j} \nu_{ij} \sigma_{ij}^2$.
Averaging over $\nu$ yields the distribution
\begin{equation}
\label{eq:Phi-z}
\Phi(z) = \int \frac{\prod_{i,j} d\nu_{ij}\, Q(\nu)}{\sqrt{2\pi \sum_{i,j} \nu_{ij} \sigma_{ij}^2}} \, e^{-z^2/(2\sum_{i,j} \nu_{ij} \sigma_{ij}^2)}.
\end{equation}
\color{black}{The limit distribution $\Phi(z)$ is therefore a mixture of Gaussian distributions of variance $\sum_{i,j} \nu_{ij} \sigma_{ij}^2$, each term in the mixture being characterized by different $\nu_{ij}$'s.
The central limit theorem then holds when $\sum_{i,j} \nu_{ij} \sigma_{ij}^2$ takes the same value for all $\nu$ having a nonzero probability weight $Q(\nu)$,
which happens} either when all $\sigma_{ij}$'s
(associated to a nonzero $\nu_{ij}$) are equal, or when the $\nu_{ij}$'s
take deterministic values, \color{black}{meaning that $Q(\nu)$ picks up a single value
of $\nu$, so that there is only one term in the mixture.}


As seen above, the distribution $Q(\nu)$ is thus the key ingredient
to characterize the limit distributions $\Psi(s)$ and (when the law of large
numbers holds) $\Phi(z)$.
In the following, we determine $Q(\nu)$ in representative cases,
yielding typical examples of limit distributions.

\section{Ergodic chain $\Gamma$}

Let us first consider the case where the chain $\Gamma$ is ergodic.
Classical Markov chain theory tells us that a homogeneous chain $\Gamma$ is ergodic if and only if its transition matrix is irreducible \cite{Seneta06}.
Moreover, \color{black}{if the structure matrix $\mE$ is irreducible and aperiodic, the Perron-Frobenius theorem \cite{Seneta06} implies that $\mE$
admits a dominant positive eigenvalue $\lambda_1$. Consequently, injecting $(\mE^n)_{ij} \approx \lambda_1^n \mu_i \nu_j$
into Eq.~(\ref{eq:transition}) shows that} the Markov chain is quasi-homogeneous far from its end point and its homogeneous transition matrix is also irreducible.
From standard results of homogeneous Markov Chain theory \cite{Seneta06},
the chain converges to a unique stationary state described by a probability
$q_i$ to occupy state $i$.
The convergence towards the stationary state implies that, for $N \to \infty$,
$\nu_{ij}$ converges (almost surely) to the average value
$\overline{\nu}_{ij} = q_i {\rm Pr}(\Gamma_{k+1}= j | \Gamma_{k}=i)$.
The empirical frequencies $\nu_{ij}$ thus do not fluctuate and both
the law of large numbers and the central limit theorem hold.
This can be seen formally by plugging the distribution
$Q(\nu) = \prod_{i,j} \delta\left(\nu_{ij}-\overline{\nu}_{ij}\right)$
into Eqs.~\eqref{eq:Psi-s} and \eqref{eq:Phi-z}.
Interestingly, this result is independent of the form of the matrix $\mA$.
Note also that the above result is consistent with the fact that
correlations, \color{black}{computed from Eq.~(\ref{eq:correlation}),} are exponentially decreasing when the matrix $\mE$ is irreducible,
\color{black}{due to the presence of a dominant eigenvalue $\lambda_1$ of multiciplity $1$:
\begin{equation}
C_{kl} \approx \sum_{l=2}^{D} K_l \left(\frac{\lambda_l}{\lambda_1}\right)^{|k-l|}
\end{equation}
with $K_l$ real constants \cite{Angeletti-TSP}.
}

\section{Chain $\Gamma$ with disconnected ergodic components}

Nonstandard distributions are expected to emerge
when the chain $\Gamma$ is non-ergodic. A potential source of non-ergodicity is 
the presence of disconnected ergodic components, that is, subdomains of configuration space between which no transitions are possible.
The simplest case appears when $\mE$ is equal to the identity matrix $I$
(a proportionality factor can be scaled out by a redefinition of
$\mR(x)$).
From Eq.~\eqref{eq:transition}, one sees that the only possible chains
are the constant chains $\Gamma^{(i)}=(i,i,\ldots,i)$, for $i=1,\ldots D$.
The chain $\Gamma$ is therefore trapped inside state $i$ and cannot explore the
whole domain.
For a given chain $\Gamma^{(i)}$, one thus has $\nu_{ii}=1$
and $\nu_{kl}=0$ for $(k,l) \neq (i,i)$.
In addition, Eq.~\eqref{eq:kappa} implies that the chain $\Gamma^{(i)}$
appears with probability $q_i = \mA_{ii}/\sum_j \mA_{jj}$.
As a result, one finds
\begin{equation}
Q(\nu) = \frac{1}{D}\sum_{i=1}^D q_i\, \delta\left(\nu_{ii}-1\right)
\left[\prod_{(k,l)\neq (i,i)} \delta \left(\nu_{kl}\right)\right],
\end{equation}
which, from Eq.~\eqref{eq:Psi-s}, leads to a generalization of the
law of large numbers,
\begin{equation}
\label{eq:Psi-s:diag}
\Psi(s) = \frac{1}{D}\sum_{i=1}^D q_i \, \delta\left(s-m_{ii}\right).
\end{equation}
In other words, in a large sample the scaled sample mean can take several
distinct values.
If all the $m_{ii}$'s are equal, the standard law of large numbers holds,
and using Eq.~\eqref{eq:Phi-z}, the central limit theorem is generalized
into
\begin{equation}
\label{eq:Phi-z:diag}
\Phi(z) = \frac{1}{D}\sum_{i=1}^D
\frac{q_i}{\sqrt{2\pi \sigma_{ii}^2}} \, e^{-z^2/2\sigma_{ii}^2}.
\end{equation}
\color{black}{Another characteristic of this form of non-ergodicity is the presence
of a non-zero constant correlation $C_{kl}$ which reads, using
Eq.~(\ref{eq:correlation}) with $\mE = I$\color{black}{,}
\begin{equation} 
C_{kl} = \frac{ \Lf(M(1)^2) -\Lf(M(1))^2}{\Lf(I)}.
\end{equation}
In terms of Markov chains, this constant correlation can be interpreted as resulting from the permanent trapping of the chain $\Gamma$ inside one of the ergodic components \cite{Angeletti-TSP}.

}

\section{Non-ergodic chain $\Gamma$ with irreversible transitions}

\begin{figure}[t]
\centering\includegraphics[width=7.5cm,clip]{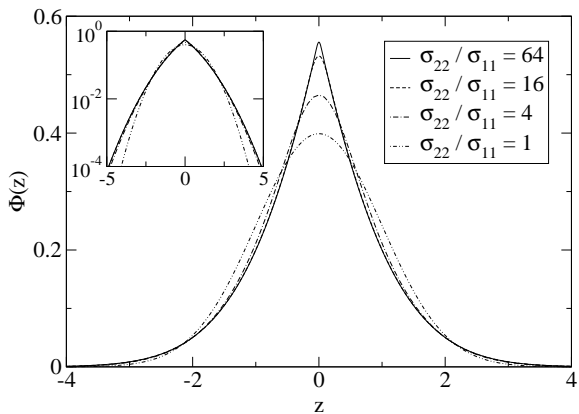}
\caption{\sl Limit distribution $\Phi(z)$ generalizing
the Gaussian distribution in the case of two-dimensional matrices
--see Eq.~\eqref{eq:Phi-z2d}, for different ratios $\sigma_{11}/\sigma_{22}$.
Parameters are chosen such that all distributions have variance $1$.
The Gaussian distribution is recovered for $\sigma_{11}=\sigma_{22}$.
Inset: same data on a semi-logarithmic scale.
}
\label{fig-phiz}
\end{figure}

Another major potential source of non-ergodicity in the chain $\Gamma$
is the presence of irreversible transitions: some subdomains $D_1,\dots,D_p$
of the configuration space of the chain
are such that the chain $\Gamma$ can never go from $D_k$ to $D_l$ if $k>l$. 
A simple and representative example of this case consists in the matrix
\begin{equation} \label{eq:smm}
\mE = I + U, \quad U=
\begin{pmatrix}
 0 &   b    &       & 0 \\
   & \ddots & \ddots&   \\
   &        & \ddots& b \\
0  &        &       & 0 \\ 

\end{pmatrix}
\end{equation}
with \color{black}{ $b$ an arbitrary strictly positive real.}
\color{black}{From Eq.~\eqref{eq:transition}, chains $\Gamma$ having nonzero probability contain only transitions from $i$ to $i$ and from $i$ to $i+1$, and thus take the form
\begin{equation}
\Gamma  = (i_1 , \dots , i_1, i_2 \dots, i_{p-1}, i_p , \dots , i_p ),
\quad i_1<i_2<\dots<i_p.     
\end{equation}
Using the form \eqref{eq:smm} of the matrix $\mE$, Eq.~\eqref{eq:kappa} implies that all the chains sharing the same start and end points are equiprobable.
Moreover, using the expansion
\begin{equation} \label{expans-E}
\mE^N = \sum_{n=0}^{D-1} b^{n} \binom{N}{n} U^n, \qquad N \ge D-1,
\end{equation}
in Eq.~\eqref{eqn:trans0}, the probability for a chain to start at $k$ and end at $l$ reads
\begin{equation}
{\rm Pr}(\Gamma_1=k, \Gamma_{N+1}=l) = \frac{b^{l-k} {N \choose {l-k}} \mA_{k,l}}{\sum_{k' \le l'} b^{l'-k'} {N \choose {l'-k'}} \mA_{k',l'}}  
\end{equation}
for $k \le l$, and zero for $k>l$.
Since ${N \choose r}$ grows as $N^r$ for $N \to \infty$,
one finds in this limit that, with probability $1$,
the chain starts from $\Gamma_1=1$ and ends at $\Gamma_{N+1}=d$,
on condition that $\mA_{1,n}$ is non-zero.}
Given that for $i \neq j$, the chains $\Gamma$ have at most
one transition from $i$ to $j$, the corresponding empirical frequencies
$\nu_{ij}$ converge to zero for $N \to \infty$.
Only the frequencies $\nu_{ii}$ are nonzero, and the equiprobability
of chains implies that they are uniformly distributed, under the
constraint $\sum_i \nu_{ii}=1$.
Hence, at odds with previous cases, empirical frequencies have
a continuous distribution in the limit $N \to \infty$,
\begin{equation}
Q(\nu) = \frac{1}{(D-1)!} \; \delta\!
\left(\sum_{i=1}^D \nu_{ii} -1\right)
\prod_{k\neq l} \delta\left( \nu_{kl}\right) \,.
\end{equation}
From Eq.~\eqref{eq:Psi-s}, the generalized law of large numbers reads
\begin{equation}
\label{eq:Psi-s:nondiag}
\Psi(s) = \int \frac{\prod_{i=1}^D d\nu_{ii}}{(D-1)!}\, \delta\!
\left(\sum_{i=1}^D \nu_{ii} -1\right)
\delta \! \left( s-\sum_{i=1}^D \nu_{ii} m_{ii} \right).
\end{equation}
The distribution $\Psi(s)$ is piecewise polynomial,
with support $[\min\{m_{ii}\},\max\{m_{ii}\}]$
(see below for an explicit example).
If all the $m_{ii}$'s are equal, the standard law of large numbers
holds, and a generalized form of the central limit theorem
is obtained from Eq.~\eqref{eq:Phi-z} as
\begin{equation}
\label{eq:Phi-z:nondiag}
\Phi(z) = \int \frac{\prod_i d\nu_{ii}}{(D-1)!}\, \delta\!
\left(\sum_i \nu_{ii} -1\right)
\frac{e^{-z^2/(2\sum_i \nu_{ii} \sigma_{ii}^2)}}{\sqrt{2\pi \sum_i \nu_{ii} \sigma_{ii}^2}}.
\end{equation}
As an example, we consider
Eqs.~\eqref{eq:Psi-s:nondiag} and \eqref{eq:Phi-z:nondiag}
in the case $D=2$.
Then, $\Psi(s)$ is simply a uniform distribution on the interval
$[\min\{m_{ii}\},\max\{m_{ii}\}]$, and if $m_{11}=m_{22}$,
\begin{equation}
\label{eq:Phi-z2d}
\Phi(z) = \sqrt{\frac{2}{\pi}}
\int_{\sigma_{11}}^{\sigma_{22}} \frac{d\sigma}{\sigma_{22}^2-\sigma_{11}^2} \;
e^{-z^2/2 \sigma^2}.
\end{equation}
This distribution is illustrated in Fig.~\ref{fig-phiz} for different values
of $\sigma_{22}/\sigma_{11}$, keeping the variance of $\Phi(z)$ fixed to $1$.
Increasing $\sigma_{22}/\sigma_{11}$ makes the central peak
of the distribution sharper, while the tails remain essentially Gaussian.

\color{black}{Another consequence of the form \eqref{eq:smm} of the matrix $\mE$ is that the correlation is long-range and becomes a function of $k/N$ and $l/N$.
Using Eq.~\eqref{eq:correlation} together with the expansion (\ref{expans-E}),
one finds in the limit $N \rightarrow +\infty$, keeping $k/N$ and $l/N$ fixed
\begin{equation}
\begin{aligned}
\langle x_k \rangle &\approx \sum_{ r+s =D-1 } a_{r,s} \left(\frac{k}{N}\right)^{r} \left(1- \frac{k}{N}\right)^{s},\\
\langle x_k x_l \rangle &\approx \sum_{ r+s+t =D-1 } c_{r,s,t} \left(\frac{k}{N}\right)^{r} \left(\frac{k-l}{N}\right)^{s} \left(1- \frac{l}{N} \right)^{t}
\end{aligned}
\end{equation}
with $a_{r,s}$ and $c_{r,s,t}$ real constants \cite{Angeletti-TSP}.
}

More generally, for an arbitrary matrix $\mE$, an analysis can be performed
in terms of decomposition of $\mE$ into irreducible blocks.
This general case can be interpreted
as a combination of the different cases above, involving partial equilibration
of the chain in some domains, irreversible transitions between domains,
and disconnected domains.
The final result for the distribution $\Psi(s)$
(or when applicable, for $\Phi(z)$) is generically a complicated mixture,
both continuous and discrete, of standard laws. Details will be
given in a forthcoming publication \cite{Angeletti-unpublished}.

\section{Discussion}

\color{black}{It is of interest to try to make a connection between the formalism presented here and nonequilibrium stochastic models which can be solved using a matrix product ansatz. An immediate difficulty is that the condition that all coefficients of matrices $\mA$ and $\mR(x)$ should be non-negative is violated in most known examples with finite matrices \cite{Evans07}. However, in some cases including for instance the ASEP model \cite{Mallick97} and the coagulation-decoagulation model \cite{Hinrichsen96}, only matrix $\mA$ contains negative coefficients, and our approach can be generalized by separating positive and negative terms in the operator $\Lf(M)$ in Eq.~(\ref{eq:Def}), writing $\Lf(M)=\Lf_{+}(M)-\Lf_{-}(M)$, which yields $P(x_1,\dots,x_N)=P_{+}(x_1,\dots,x_N)-P_{-}(x_1,\dots,x_N)$.
The Markov chain reformulation can then be applied to each part $P_{+}$ and $P_{-}$ separately (after proper normalization), and the distribution of the sum $S=\sum_{k=1}^N x_k$ can be recomposed from the two distributions of $S$ obtained from $P_{+}$ and $P_{-}$.

In addition, let us discuss a tentative physical interpretation of the chain $\Gamma$, in the case of a non-ergodic chain with irreversible transitions.} Considering for instance two-dimensional matrices,
a typical chain has the form $\Gamma=(1,\ldots,1,2,\ldots,2)$,
with a single transition from state $1$ to state $2$.
From Eq.~(\ref{eq:X|gamma}), the probability distribution
$P_{\Gamma} \equiv P_{\Gamma}(x_1,\ldots,x_N)$ associated to $\Gamma$ reads
\begin{equation} \label{eq:shock}
P_{\Gamma} = \mP_{11}(x_1) \ldots \mP_{11}(x_k)
\mP_{22}(x_{k+1}) \ldots \mP_{22}(x_N),
\end{equation}
and precisely corresponds to a Bernoulli shock measure,
known to be the building block of the dynamics in the ASEP
and in lattice reaction-diffusion models \cite{Krebs03}
($x_k=0$ or $1$ characterizes the occupancy of site $k$).
These shocks are not purely formal objects, but rather describe the
typically observed configurations on the coexistence line of the ASEP
\cite{Schutz93,Appert02} and related models \cite{Jafarpour03}.
The fact that the position of the shock is uniformly distributed
over the system \cite{Schutz93,Depken04}
reflects the random transition in the chain $\Gamma$,
and results in a uniform distribution $\Psi(s)$ on an interval
$[\rho_{\min},\rho_{\max}]$, of the scaled number of particles.
Such models thus provide explicit realizations of systems where the scaled
sample mean can take several values (here a continuum of values),
thus breaking the law of large numbers.
\color{black}{Note that the observation of breakings of the central limit theorem (in situations where the law of large numbers is valid) cannot be observed
in such models with binary variables $x_k$, where the random variables $x_k$ take only two values $0$ and $1$. This would require that all first moments $m_{ii}$ are equal, which for binary variables implies that the distributions $\mP_{ii}(x)$ are identical.
The introduction of models with continuous variables \cite{Bertin-unpublished}
should thus be helpful in identifying physical situations where the generalized
central limit theorem obtained in Eq.~(\ref{eq:Phi-z:nondiag}) is valid.}

\color{black}{In conclusion, we have derived generalizations of the law of large numbers
and of the central limit theorem for strongly correlated variables
described by a matrix product ansatz,
a class of random variables ubiquitous in the description
of one-dimensional non-equilibrium systems.
We have shown that the type of distribution found is related both to the ergodicity properties of the associated Markov chain, and to the short-range or long-range nature of the correlation.}
We believe that beyond their applicability to exactly
solvable models, our results also provide families of reference distributions,
like the one given in Eq.~(\ref{eq:Phi-z2d}),
which can be useful to describe experimental or numerical data
in more general correlated non-equilibrium systems.
\color{black}{As for future work, it would be interesting to try to extend these results to matrices with coefficients of arbitrary signs, as well as to infinite matrices. Providing a more physical interpretation of the Markov chain in terms of dynamics of the underlying stochastic model would also be valuable.}


\bibliographystyle{eplbib}

\bibliography{SumStat}

\end{document}